\documentclass[aps,showpacs]{revtex4}
\usepackage{amsmath}
\usepackage{graphicx}
\begin{document}

\begin{flushright}
\fbox{UCRL-TR-200505}
\end{flushright}

\title{Nuclear structure with accurate chiral perturbation 
theory nucleon-nucleon potential: Application to $^6$Li and $^{10}$B}
\author{P. Navr\'atil}
\affiliation{Lawrence Livermore National Laboratory, L-414, P.O. Box 808, 
Livermore, CA  94551}
\author{E. Caurier}
\affiliation{Institut de Recherches Subatomiques
            (IN2P3-CNRS-Universit\'e Louis Pasteur)\\
            Batiment 27/1,
            67037 Strasbourg Cedex 2, France}

\begin{abstract}
We calculate properties of $A=6$ system using the accurate charge-dependent
nucleon-nucleon (NN) potential at fourth order of chiral perturbation theory.
By application of the {\it ab initio} no-core shell model (NCSM) 
and a variational
calculation in the harmonic oscillator basis with basis size up to $16\hbar\Omega$
we obtain the $^6$Li binding energy of 28.5(5) MeV and a converged excitation
spectrum. Also, we calculate properties of $^{10}$B using the same NN potential in 
a basis space of up to $8\hbar\Omega$. Our results are consistent 
with results obtained by standard
accurate NN potentials and demonstrate a deficiency of Hamiltonians consisting
of only two-body terms. At this order of chiral perturbation theory three-body
terms appear. It is expected that inclusion of such terms in the Hamiltonian
will improve agreement with experiment. 
\end{abstract}
\maketitle

\section{Introduction}

A new development in the theory of nuclear forces occurred when the concept
of an effective field theory (EFT) was introduced and applied to the low-energy
QCD \cite{Weinberg}. Starting from a general Lagrangian consistent with the
QCD symmetries, in particular the broken chiral symmetry, one can develop
a systematic perturbative expansion valid at low energies for effective degrees 
of freedom, nucleons and pions. Pioneering work in this direction was performed 
by Ordonez, Ray and van Kolck \cite{ORK94,Bira} who constructed a nucleon-nucleon 
(NN) potential in the coordinate space based on the chiral perturbation 
theory at the third order.
Similarly, Epelbaum {\it et al.} \cite{Epelbaum} developed the first momentum space NN 
potential at the third order of the chiral perturbation theory. Recently, Entem and 
Machleidt developed an EFT based momentum space non-local NN potential at the fourth
order of the chiral perturbation theory (next-to-next-to-next-to-leading order, N$^3$LO) 
\cite{N3LO} that takes into account
the charge dependence and fits the two-nucleon data with similar accuracy as the
traditional phenomenological high-quality NN potentials like Argonne V18 \cite{av18} or 
the CD-Bonn 2000 \cite{cdb2k}.   

It is an important question and a test for the EFT approach how well these new potentials
will describe the nuclear structure of light nuclei when applied to more than two nucleons.
The new accurate N$^3$LO NN potential is a non-local momentum-space potential. There are
several methods that could be applied to solve $A=3,4$ systems using this potential,
most notably the Faddeev and the Faddeev-Yakubovsky techniques \cite{FY}. However, 
the best established method for obtaining exact results for the $A>4$ $p$-shell nuclei,
the Green's Function Monte Carlo (GFMC) approach \cite{wiringa00,pieper01,GFMC_9_10}, 
is not applicable as it is best suited for work with local coordinate space potentials like 
AV18 or AV8$^\prime$ \cite{av8p}. The N$^3$LO potential can be used, however, in the
{\it ab initio} no-core shell model approach (NCSM) \cite{C12_NCSM}. In this method, we
perform calculations in a finite harmonic-oscillator (HO) basis using effective interactions
appropriate to the basis size truncation that are systematically derived from the original
inter-nucleon potential. The method is convergent to the exact solution with the basis size 
enlargement and/or with the degree of clustering of the effective interaction. Due to the 
properties of the HO basis it is no technical problem to use either local or non-local and
either coordinate or momentum space potentials.

In this paper, we apply the NCSM to obtain nuclear structure results using the N$^3$LO NN
potential for $^3$H, $^4$He, $^6$Li, $^6$He and $^{10}$B. With the improved shell model code 
ANTOINE \cite{Antoine,Be8_NCSM} we are able to reach basis sizes up to the $16\hbar\Omega$ 
for $^6$Li
and up to the $8\hbar\Omega$ for $^{10}$B. At the same time, we perform variational calculations
in the HO basis for $^6$Li using the bare N$^3$LO potential. These variational calculations 
allow us to make extrapolations and check our NCSM effective interaction results. 
In addition, we compare the N$^3$LO results to those obtained by the CD-Bonn 2000 
and the AV8$^\prime$.   

It should be pointed out that already at the third order of the chiral perturbation
theory, N$^2$LO, three-body terms appear \cite{EFT_V3b}. These terms should be
consistently included in calculations for $A>2$ nuclei. It is our goal to  
perform the NCSM calculations with the three-body terms included using the approach 
outlined in Ref. \cite{NO03}. In fact, work in this direction is under way \cite{Andreas}.
However, in general when only the two-nucleon interactions are employed the NCSM
approach can achieve a higher accuracy as larger basis spaces can be reached. Therefore,
it is valuable to perform accurate calculations using just the two-nucleon forces
to assess the effects of the omitted three-nucleon terms. The eventual less accurate 
results obtained with both the two- and three-nucleon forces can then be better 
understood and extrapolated.    

We briefly review the basic features of the NCSM and describe our variational calculations 
in the HO basis in Section \ref{sec_theory}.
In Section \ref{sec_test}, we present our N$^3$LO results for $^3$H and $^4$He.
In Section \ref{sec_res_a=6}, we discuss our calculations for $A=6$ using the N$^3$LO
NN potential and compare them to results obtained by the AV8$^\prime$ and the CD-Bonn 2000.
In addition, we check convergence properties of our approach using the semi-realistic
Minnesota NN potential. In Section \ref{sec_res_b10}, we show the NCSM predictions for 
$^{10}$B with the N$^3$LO NN potential. Our conclusions are summarized in Section 
\ref{sec_concl}.  
 
\section{{\it Ab initio} no-core shell model 
and variational calculations in the HO basis}\label{sec_theory}

A detailed description of the NCSM approach was presented, e.g. 
in Refs. \cite{C12_NCSM,fourb_NCSM,Jacobi_NCSM}. Here we only briefly review 
the basic features of the NCSM. Also, we describe our variational calculations 
in the HO basis that we apply in the following Sections.

The starting Hamiltonian for our investigations is
\begin{equation}\label{ham}
H_A= 
\frac{1}{A}\sum_{i<j}^{A}\frac{(\vec{p}_i-\vec{p}_j)^2}{2m}
+ \sum_{i<j}^A V_{{\rm NN}, ij}  \; ,
\end{equation}
where $m$ is the nucleon mass, $V_{{\rm NN}, ij}$,
the NN interaction with both strong and electromagnetic components. 
In the NCSM, we employ a large
but finite harmonic-oscillator (HO) basis. Due to properties of the realistic nuclear 
interaction in Eq. (\ref{ham}),
we must derive an effective interaction appropriate for the basis truncation
in order to reach convergence or to achieve a reasonable approximation 
to the exact solution.
To facilitate the derivation of the effective interaction, we modify the
Hamiltonian (\ref{ham}) by adding to it the center-of-mass (CM) HO Hamiltonian
$H_{\rm CM}=T_{\rm CM}+ U_{\rm CM}$, where
$U_{\rm CM}=\frac{1}{2}Am\Omega^2 \vec{R}^2$,
$\vec{R}=\frac{1}{A}\sum_{i=1}^{A}\vec{r}_i$.
This modification was first introduced by Lipkin \cite{Lipkin}.
The effect of the HO CM Hamiltonian will later be subtracted
out in the final many-body calculation. Due to the translational invariance of the
Hamiltonian (\ref{ham}) the HO CM Hamiltonian has in fact no effect on the intrinsic
properties of the system in the infinite basis space or in a finite basis space truncated
by $N_{\rm max}$ as described below if the interaction in Eq. (\ref{ham}) is not altered. 

The modified Hamiltonian can be cast into the form
\begin{eqnarray}\label{hamomega}
H_A^\Omega &=& H_A + H_{\rm CM}=\sum_{i=1}^A h_i + \sum_{i<j}^A V_{ij}^{\Omega,A}
 \nonumber \\ 
&=& \sum_{i=1}^A \left[ \frac{\vec{p}_i^2}{2m}
+\frac{1}{2}m\Omega^2 \vec{r}^2_i
\right] + \sum_{i<j}^A \left[ V_{{\rm NN}, ij}
-\frac{m\Omega^2}{2A}
(\vec{r}_i-\vec{r}_j)^2
\right] \; .
\end{eqnarray}
Next we divide the $A$-nucleon infinite HO basis space
into the finite active space ($P$) comprising of all states of up to $N_{\rm max}$
HO excitations above the unperturbed ground state and the excluded space ($Q=1-P$).  
The basic idea of the NCSM approach is to apply a unitary transformation
on the Hamiltonian (\ref{hamomega}), $e^{-S} H_A^\Omega e^S$ such that
$Q e^{-S} H_A^\Omega e^S P=0$ \cite{LS1,LS2}. If such a transformation is found, the effective
Hamiltonian that exactly reproduces a subset of eigenstates of the full space Hamiltonian
is given by $H_{\rm eff}=P e^{-S} H_A^\Omega e^S P$. This effective Hamiltonian
contains up to $A$-body terms and to construct it is essentially as difficult as to solve
the full problem. Therefore, we apply this basic idea on a sub-cluster level.
In this paper we use the simplest approximation, a two-body effective interaction approximation.
This approximation is obtained by constructing the unitary transformation for the Hamiltonian
(\ref{hamomega}) applied to two nucleons only, i.e., ${\cal H}_2=h_1+h_2+V_{12}^{\Omega,A}$
with $A$ in $V_{12}^{\Omega,A}$ fixed at the value corresponding to the final
$A$-nucleon calculation. The two-body effective interaction is then derived with the help
of the exact eigensolutions of ${\cal H}_2$ as 
$[V_{12}^{\Omega,A}]_{\rm eff}=P_2 \left[e^{-S_2} {\cal H}_2 e^S_2 -h_1-h_2 \right] P_2$. 
Here, the definition of the two-nucleon model space projector $P_2$ follows directly
from the definition of the $A$-nucleon projector $P$. The two-nucleon transformation $S_2$
is determined by the condition $Q_2 e^{-S_2} {\cal H}_2 e^{S_2} P_2=0$. 

The $A$-nucleon calculation with $A>4$ is then
performed using the multi-shell (no-core) version of the code ANTOINE,
with the Hamiltonian in the form: 
\begin{equation}\label{Ham_A_Omega_eff_SM}
H^{\Omega, {\rm SM}}_{A, {\rm eff}}=P \left\{ \sum_{i<j}^A \left[ \frac{(\vec{p}_i-\vec{p}_j)^2}{2Am}
+\frac{m\Omega^2}{2A} (\vec{r}_i-\vec{r}_j)^2\right]
+ \sum_{i<j}^A 
\left[ V_{{\rm NN}, ij}-\frac{m\Omega^2}{2A} (\vec{r}_i-\vec{r}_j)^2 \right]_{\rm eff}
+ \beta(H_{\rm CM}-\frac{3}{2}\hbar\Omega)\right\} P \; ,
\end{equation}
where we subtracted the CM Hamiltonian $H_{\rm CM}$ and added the Lawson projection term 
$\beta(H_{\rm CM}-\frac{3}{2}\hbar\Omega)$ to shift the spurious
CM excitations. As our effective interaction is translationally invariant our physical
eigenenergies do not depend on the choice of $\beta$. Note that in Section \ref{sec_test}
we present $A=3,4$ calculations performed in Jacobi coordinate antisymmetrized HO basis. 
In those calculations the CM degrees of freedom are explicitly omitted and therefore
no $\beta(H_{\rm CM}-\frac{3}{2}\hbar\Omega)$ term is used. 

We note that our effective interaction depends on the nucleon
number $A$, the HO frequency $\Omega$, and the $P$-space basis size defined by 
$N_{\rm max}$. It is 
translationally invariant and with $N_{\rm max}\rightarrow \infty$ the NCSM effective
Hamiltonian approaches the starting bare Hamiltonian (\ref{ham}). Consequently, with the basis
size increase the NCSM results become less and less dependent on the HO frequency
and converge to the exact solution. Alternatively, by increasing the clustering
of the effective interaction for a fixed $P$-space size the NCSM effective Hamiltonian
approaches the exact $A$-nucleon effective Hamiltonian that reproduces exactly a subset of
eigenstates of the starting Hamiltonian (\ref{ham}).

It turns out that the N$^3$LO NN potential is softer than the phenomenological high-quality
NN potentials like the AV18 or the CD-Bonn. As we are able to reach basis spaces of up
to $N_{\rm max}=18$ and $N_{\rm max}=16$ for $A=4$ and $A=6$, respectively, it make sense
to perform variational calculations in the HO basis for $^4$He and $^6$Li using the
bare N$^3$LO NN potential with the HO frequency or the oscillator length as the only 
variational parameter. In this case, the Hamiltonian that we use is $P H_A P$, with $H_A$ given in 
Eq. (\ref{ham}) (plus the $\beta P(H_{\rm CM}-\frac{3}{2}\hbar\Omega)P$ term 
when Slater determinant basis is utilized, i.e., for $A>4$). We perform variational
calculations for different basis sizes defined by $N_{\rm max}$. We then extrapolate
to $N_{\rm max}\rightarrow \infty$ and compare to our effective interaction results.

\section{$^3$H and $^4$He results}\label{sec_test}

As a test of convergence of our method and also as a test of correct interfacing
of our codes with the N$^3$LO NN potential code \cite{Machl_priv} we performed calculations
for $^3$H and $^4$He. We used the translationally invariant HO basis antisymmetrized
as described in Ref. \cite{Jacobi_NCSM} and employed the many-body effective interaction
code MANYEFF \cite{Jacobi_NCSM}.
Our results are summarized in Table \ref{tab_he4}. 
Our $^3$H binding energy, 7.85(1) MeV, that was obtained using basis spaces 
up to $N_{\rm max}=40$ agrees
well with the result obtained by the Faddeev method \cite{Machl_priv}. 

\begin{figure}
\vspace*{2cm}
\includegraphics[width=7.0in]{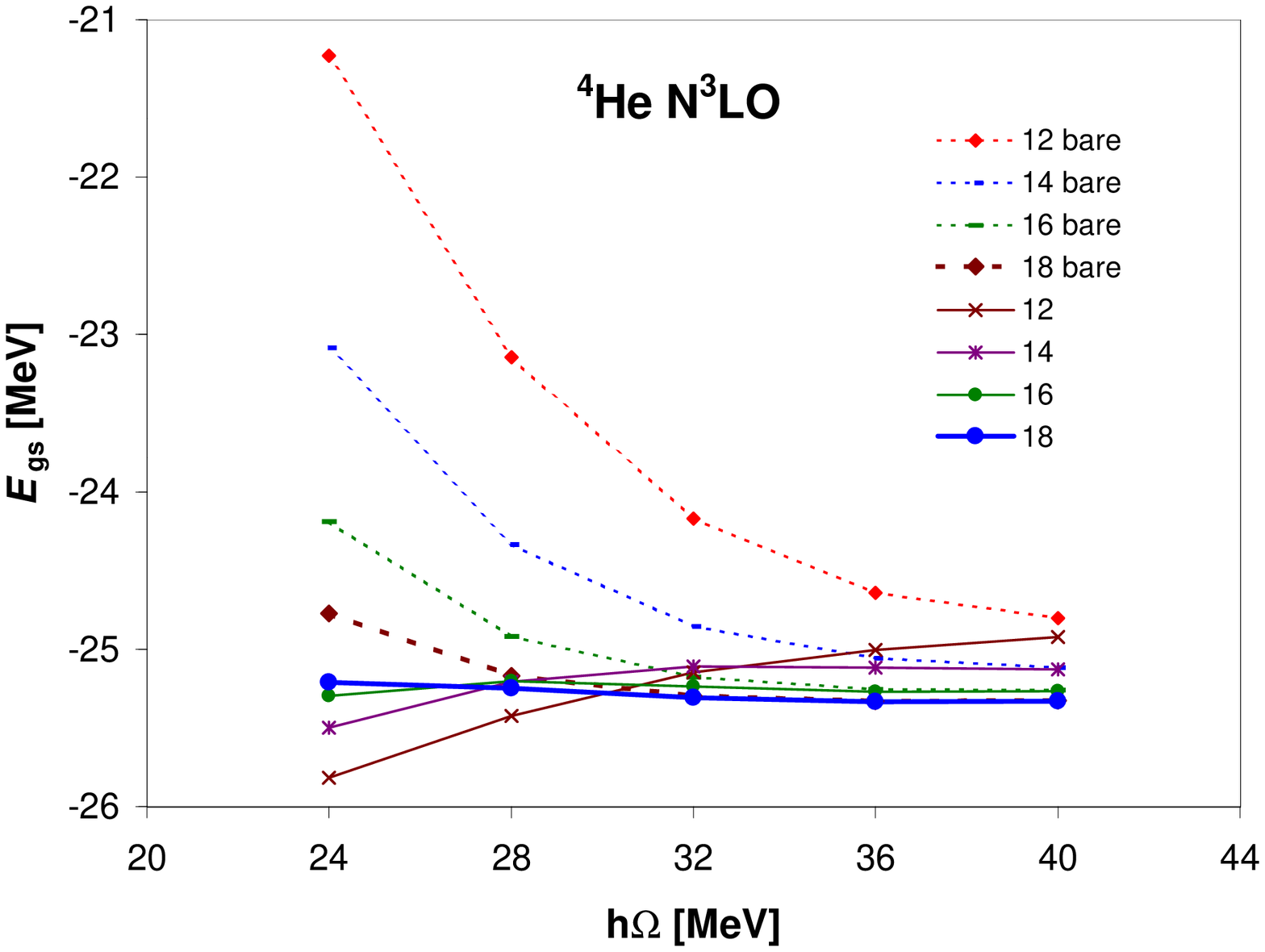}
\caption{\label{he4_n3lo} (Color online) Harmonic-oscillator frequency dependence 
of the $^{4}$He ground-state energy 
obtained in $12\hbar\Omega$-$18\hbar\Omega$ ($N_{\rm max}=12-18$) basis spaces using 
the bare N$^3$LO NN potential (dotted lines) and two-body effective interactions 
derived from the N$^3$LO NN potential (full lines). The Coulomb potential is included. 
}
\end{figure}

Our $^4$He
binding energy calculations were performed using both bare and the two-body effective
interactions in basis spaces up to $N_{\rm max}=18$. Our obtained binding energy, 25.36(4) MeV,
is very close to that obtained by the Faddeev-Yakubovsky method \cite{Andreas}.  
In Fig. \ref{he4_n3lo}, we present the HO frequency dependence of our $^4$He binding energy 
for different basis sizes. Calculations with two-body effective interactions
are compared to the variational calculations using the bare N$^3$LO NN potential. 
The NCSM calculations with the two-body effective interaction are not variational
as some terms of the transformed Hamiltonian (i.e., the higher than two-body terms) are omitted.
Therefore, in the NCSM approach the convergence could be, depending on the HO frequency, 
either from above or below or even oscillatory with the basis size enlargement.
It should be noted how the frequency dependence of our $^4$He binding energy 
decreases with the basis size enlargement. Also, the two-body effective 
interaction always improves on the bare interaction results in particular for the smaller spaces
and lower HO frequencies.
For example, the $N_{\rm max}=14$ ($14\hbar\Omega$) binding energy obtained using the effective
interaction varies within 25.11 MeV and 25.5 MeV in the range of HO frequencies shown 
in Fig. \ref{he4_n3lo}, $\hbar\Omega=24-40$ MeV. The bare interaction binding energy
in the same basis space and the same frequency range varies within 23.09 MeV and 25.12 MeV.
For the highest frequency that we used here, $\hbar\Omega=40$ MeV, we observe that the
improvements due to the two-body effective interaction become very small.
We note that the the minimum of the variational calculation in the $N_{\rm max}=18$ basis
space is at about $\hbar\Omega=36$ MeV. The minima for the smaller space calculations
are at still higher frequencies, e.g., the minimum of the of the $N_{\rm max}=12$ calculation
is at $\hbar\Omega>40$ MeV outside of the range shown in Fig. \ref{he4_n3lo}.
At $N_{\rm max}=18$ our effective interaction calculations agree with the variational 
bare interaction result. 

\begin{figure}
\vspace*{2cm}
\includegraphics[width=7.0in]{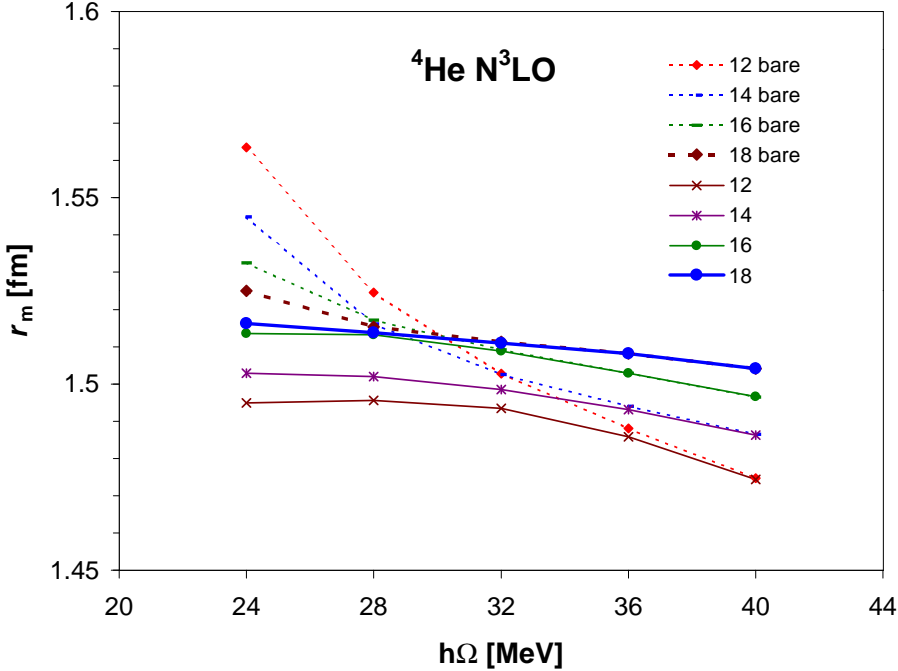}
\caption{\label{he4_n3lo_rad} (Color online)
The same as in Fig. \protect\ref{he4_n3lo} for the point-nucleon radius.
}
\end{figure}

\begin{figure}
\vspace*{2cm}
\includegraphics[width=7.0in]{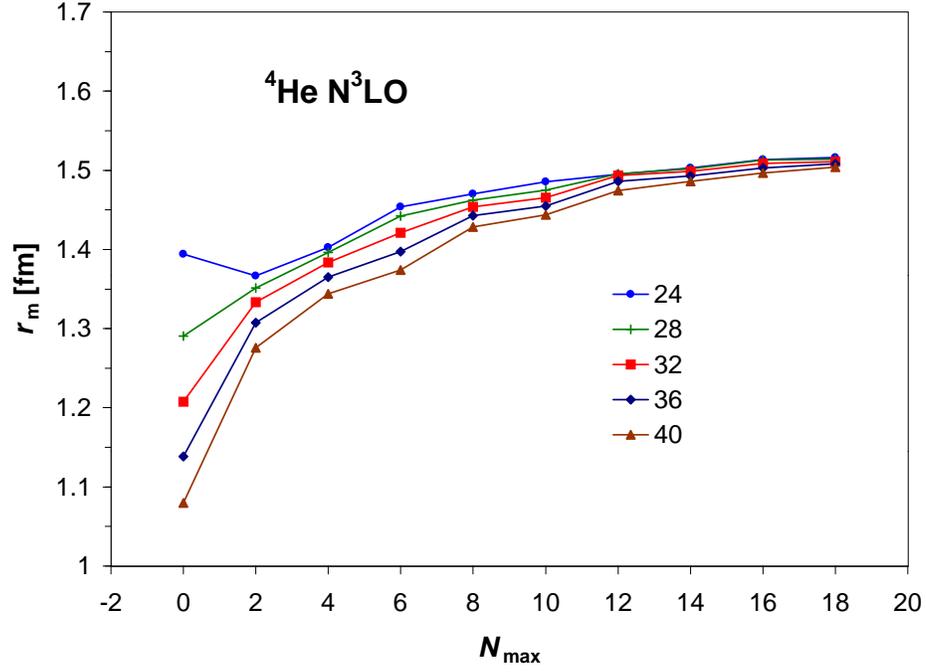}
\caption{\label{he4_n3lo_rad_nmax} (Color online)
Basis size dependence of the $^{4}$He point-nucleon radius
obtained using the two-body effective interactions 
derived from the N$^3$LO NN potential. 
Results for the harmonic-oscillator frequencies  
$\hbar\Omega=24,28,32,36,40$ MeV are presented.
}
\end{figure}

It is instructive to investigate in addition to the binding energy also the convergence of the
point-nucleon $^4$He radius. A similar plot as in Fig. \ref{he4_n3lo} is presented
for the point-nucleon root mean square radius in Fig. \ref{he4_n3lo_rad}. 
Even though no operator renormalization was used for the radius operator very similar
conclusions can be drawn as for the binding energy. The calculation with the effective interaction
(i.e., with wave functions obtained using the effective interactions)
improves convergence and reduces the frequency dependence 
compared to a calculation with the bare interaction wave functions using the same basis size.
With the basis enlargement the frequency dependence decreases. At the same time,
it should be noticed that the optimal frequency for the radius convergence is not 
necessarily the same as that for the binding energy. It is important to investigate
the frequency dependence of energies and other observables to determine the extrapolated values 
and errors. We extrapolate the $^4$He radius with the N$^3$LO potential to be 1.515(10) fm.
In Fig. \ref{he4_n3lo_rad_nmax} we show the complete basis size dependence of the point-nucleon
radius calculation using the effective interaction for several HO frequencies. 
The convergence with $N_{\rm max}$ and the decrease of frequency dependence can be clearly seen.

The $^3$H and $^4$He N$^3$LO binding energies (7.85 MeV, 25.36 MeV)  
are in between the Argonne V8$^\prime$ (7.77 MeV, 25.2 MeV)
and the CD-Bonn binding energies (8.00 MeV, 26.3 MeV) but considerably closer
to the AV8$^\prime$ results. Note that the Coulomb interaction is included for all
three potentials.

\section{Application to $^6$Li and $^6$He}\label{sec_res_a=6}

As mentioned earlier, our $A=6$ calculations reported here were obtained
using the improved multi-shell (no-core) version of the code ANTOINE
capable to reach basis spaces up to $N_{\rm max}=16$ ($16\hbar\Omega$) for
$^6$Li with the M-scheme basis dimension equal to $8\times 10^8$.
A specific challenge of the no-core calculations compared to the traditional $0\hbar\Omega$
valence nucleon calculations is the enormous number of single particle states
(e.g., there are 171 $nlj$ levels and 2280 $nljm$ states for each protons and neutrons 
in the $16\hbar\Omega$ $^6$Li calculations). This implies a huge number of operators
that need to be stored in memory. The other challenge is the huge dimension of the matrix
to be diagonalized. Both these challenges have been addressed in the improved
ANTOINE code. In particular, an efficient way of operator storage was found and a partitioning
of Lanczos vectors was introduced that eliminates the need to store the full Lanczos
vectors in memory.

\begin{figure}
\vspace*{2cm}
\includegraphics[width=7.0in]{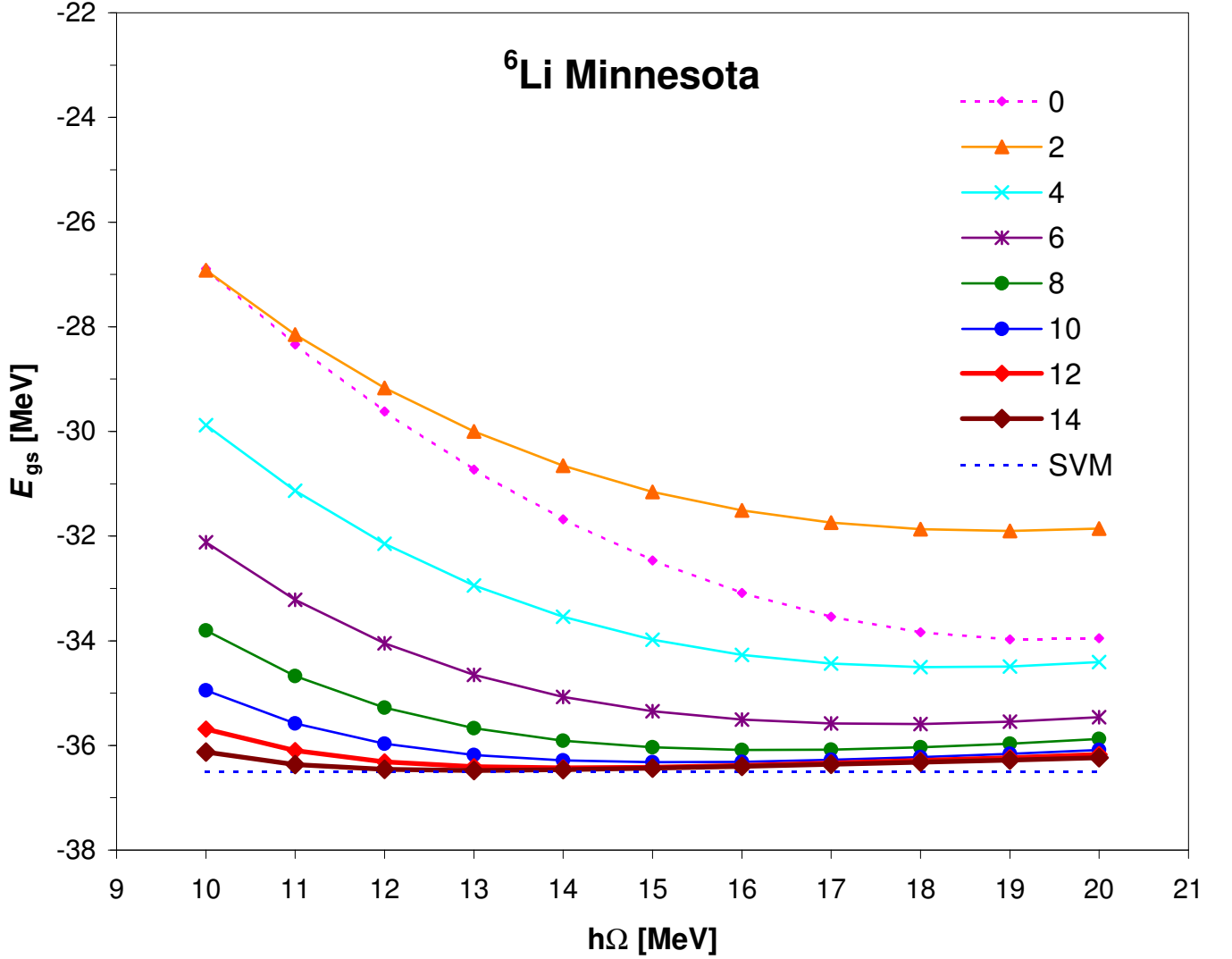}
\caption{\label{li6_mn gs_2eff} (Color online) 
Harmonic-oscillator frequency dependence of the $^{6}$Li 
ground-state energy obtained in $0\hbar\Omega$-$14\hbar\Omega$ 
basis spaces using two-body effective interactions derived from
the Minnesota NN potential. Coulomb potential is not included.
The dotted line corresponds to the Stochastic Variational Method 
result \protect\cite{Varga}.    
}
\end{figure}

Before presenting our N$^3$LO results, let us first discuss test
calculations for $^6$Li using the semi-realistic Minnesota (MN) NN potential
\cite{Minnesota} frequently used for few-body calculation benchmarks \cite{SVM,EIHH_3eff}.
Our NCSM results, in particular the HO frequency dependence of the ground-state energy
for different basis spaces ranging from $N_{\rm max}=0$ to $N_{\rm max}=14$, 
are presented in Fig. \ref{li6_mn gs_2eff}. Clearly, for the Minnesota NN potential
we observe the best case scenario of the NCSM convergence. Starting from
$N_{\rm max}=2$ the convergence is uniform in the whole frequency range. The HO frequency 
dependence is getting weaker with the basis size enlargement. The differences between 
the successive curves decrease as the basis size increases. At $N_{\rm max}=14$ the NCSM 
energy at the minimum is within 30 keV of the Stochastic Variational Method (SVM) \cite{SVM} 
result 36.51 MeV \cite{Varga}. An $N_{\rm max}=16$ calculation improves the NCSM and SVM
agreement to 10 keV. Similarly, the Effective Interaction in the Hyperspherical Harmonics 
basis (EIHH) method \cite{EIHH_3eff,EIHH} obtains a very close
binding energy of 36.64(7) MeV \cite{Winfried}.
Let us mention that a variational calculation in the HO basis using the bare Minnesota
NN potential still misses about 1.5 MeV  
from the converged value in the $N_{\rm max}=14$ basis space.
The improvement due to the two-body effective interaction is quite dramatic
even for such a large basis space.
The Minnesota NN potential was employed as a test potential for the NCSM $^6$Li investigation
already in Ref. \cite{NCSM_6}. There, however, basis spaces up to only $N_{\rm max}=10$
were utilized. Also, the Minnesota NN potential results presented in Ref. \cite{NCSM_6}
include the Coulomb interaction which was omitted in the present calculations. 
 
\begin{figure}
\vspace*{2cm}
\includegraphics[width=7.0in]{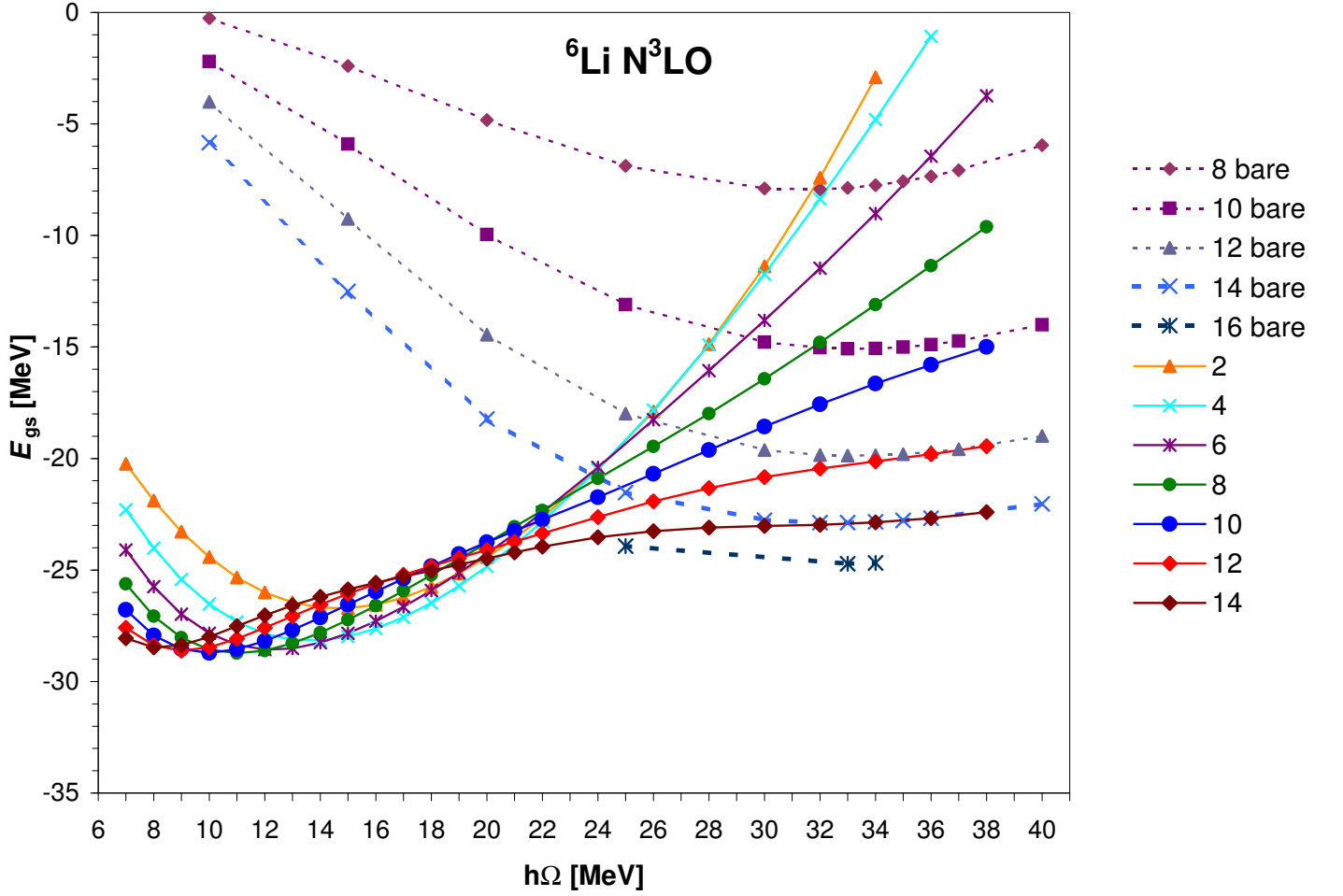}
\caption{\label{li6_gs_2eff_bare} (Color online)
Harmonic-oscillator frequency dependence 
of the $^{6}$Li ground-state energy 
obtained in $2\hbar\Omega$-$14\hbar\Omega$ and $8\hbar\Omega$-$16\hbar\Omega$ 
basis spaces using 
two-body effective interactions derived from
the N$^3$LO NN potential (full lines) and the bare N$^3$LO NN 
potential (dotted lines), respectively. 
The Coulomb potential is included.   
}
\end{figure}

\begin{figure}
\vspace*{2cm}
\includegraphics[width=7.0in]{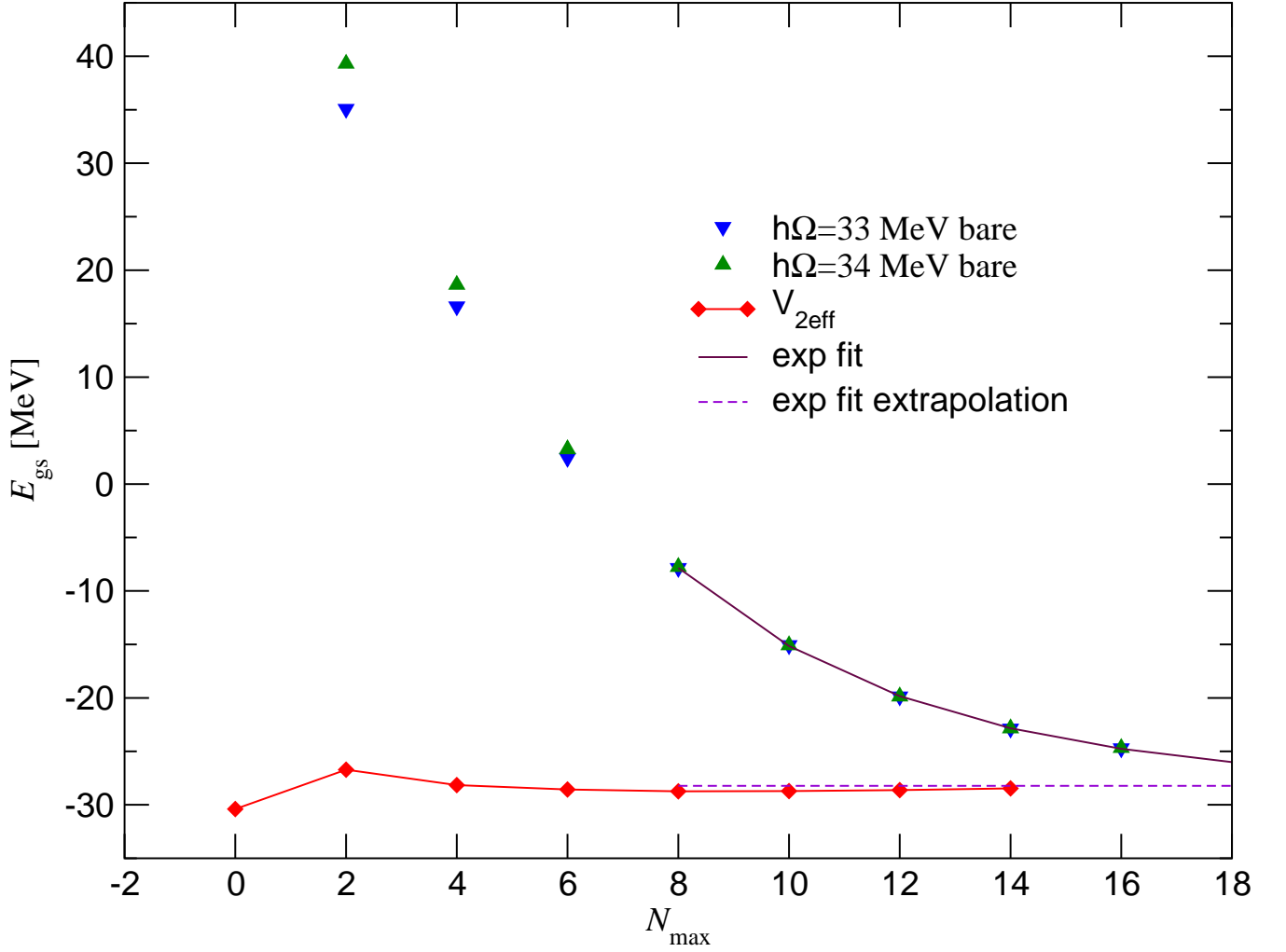}
\caption{\label{li6_n3lo_extr} (Color online)
Basis size dependence of the $^{6}$Li ground-state energy 
obtained in $0\hbar\Omega$-$14\hbar\Omega$ and $0\hbar\Omega$-$16\hbar\Omega$ 
basis spaces using 
two-body effective interactions derived from
the N$^3$LO NN potential and the bare N$^3$LO NN potential, respectively. 
The Coulomb potential is included. The two-body effective interaction
results correspond to the ground-state energy minimum for each model space.
For details on the extrapolation see the text.  
}
\end{figure}

We now turn to our $^6$Li N$^3$LO results. In Fig. \ref{li6_gs_2eff_bare}, we present
the HO frequency dependence of the ground-state energy obtained by the variational
calculations (dashed lines) in $N_{\rm max}=8$ to $N_{\rm max}=16$ basis spaces
and by the two-body effective interaction NCSM calculations in basis spaces
ranging from $N_{\rm max}=2$ to $N_{\rm max}=14$. The variational calculations
exhibit large changes with $N_{\rm max}$ and clearly they are not converged at 
$N_{\rm max}=16$. The minimum in the largest spaces that we used develops at about 
$\hbar\Omega=33$ MeV and the value at minimum for the $N_{\rm max}=16$ space is -24.7 MeV. 
The NCSM two-body effective interaction results (full lines in Fig. \ref{li6_gs_2eff_bare})
show stronger HO frequency dependence than those obtained using the Minnesota NN potential
and in fact also than those obtained using the CD-Bonn NN potential \cite{NCSM_6}.
We can see that the minimum appears around $\hbar\Omega=12$ MeV and shifts with increasing
$N_{\rm max}$ to lower frequency. Due to this shift the successive curves intersect
and we do not have a nice uniform picture as in the case of the Minnesota interaction.
At the same time, the value at minimum does not change much starting at $N_{\rm max}=6$
as can be seen in Fig. \ref{li6_n3lo_extr}
where we plot the ground-state energy obtained at the minimum for each $N_{\rm max}$.
It is interesting to note that for the largest spaces, e.g. $N_{\rm max}=14$,
a second minimum develops at around the frequency where the bare interaction
results are at the minimum. Similarly as in the case of $^4$He, we can see that starting
at certain frequency, the two-body effective interaction does not change the bare interaction
results any more and the effective interaction curves start to follow the bare interaction 
curves.

In  Fig. \ref{li6_n3lo_extr}, we also present the bare interaction variational calculation
results at fixed HO frequencies at or close to the minimum of the $N_{\rm max}=14$ 
and $N_{\rm max}=16$ spaces. It turns out that these ground-state energy $N_{\rm max}$ 
dependencies can be fitted by an exponential function of the type 
$E(N_{\rm max})=E_\infty+a e^{-b N_{\rm max}}$
(see an analogous fit done for the hyperspherical harmonics basis calculation 
in Ref. \cite{EIHH_3eff}). When we fit this formula to the last four points, $N_{\rm max}=8-16$,
we obtain $E_\infty \approx -28.2$ MeV. This is rather close to the energy
at the minima of our two-body effective interaction calculations that are at about
28.5 MeV. Based on the variational bare interaction calculations and extrapolation and on the
two-body effective interaction NCSM calculations in different spaces 
and using different frequencies
we arrive at our final $^6$Li N$^3$LO binding energy result of 28.5(5) MeV. The experimental
$^6$Li binding energy is 31.995 MeV. Our calculations suggest that the N$^3$LO NN potential
under-binds $^6$Li by about 3.5 MeV.   

\begin{figure}
\vspace*{2cm}
\includegraphics[width=7.0in]{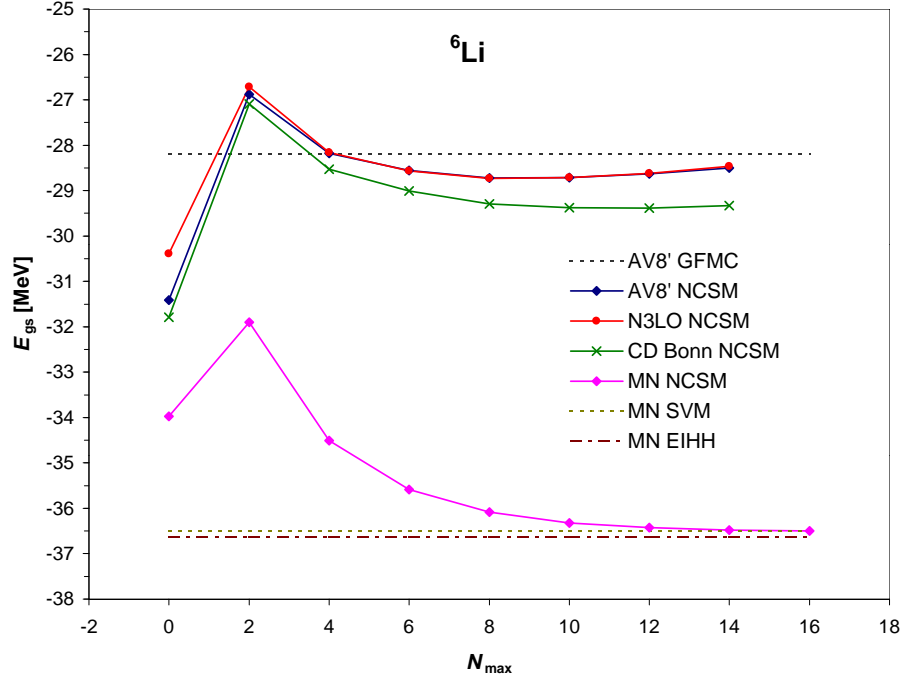}
\caption{\label{li6_mn_cdb_n3lo_v8p} (Color online) 
Basis size dependence of the $^{6}$Li ground-state energy 
obtained in $0\hbar\Omega$-$14\hbar\Omega$ ($16\hbar\Omega$ for MN)
basis spaces using 
two-body effective interactions derived from
the AV8$^\prime$, N$^3$LO, CD-Bonn 2000 and the Minnesota NN potentials. 
The AV8$^\prime$ result obtained by the GFMC method \protect\cite{pieper01} 
and the Minnesota results obtained by the SVM \protect\cite{Varga} 
and the EIHH methods \protect\cite{Winfried} are shown for a comparison.
}
\end{figure}

It is interesting to compare the N$^3$LO results to those obtained by other NN potentials.
Concerning the binding energy, in Fig. \ref{li6_mn_cdb_n3lo_v8p} we repeat 
the basis size dependence from Fig. \ref{li6_n3lo_extr}
of the NCSM ground-state energy at the frequency minima for the N$^3$LO and show the same for
the AV8$^\prime$, CD-Bonn 2000 and the Minnesota NN potentials. In addition, the AV8$^\prime$
GFMC result \cite{pieper01}  and the Minnesota SVM \cite{Varga} and EIHH \cite{Winfried} 
results are shown for comparison. As discussed earlier, our Minnesota results show a nice
convergence and agree quite well in particular with the SVM result. Our AV8$^\prime$ binding energy
is close to that obtained using the N$^3$LO NN potential and within about 300 keV of the GFMC result.
Our CD-Bonn 2000 binding energy is, on the other hand, larger by almost 1 MeV compared
to the N$^3$LO and the AV8$^\prime$. Our present result is almost the same as that presented
in Ref. \cite{NCSM_6} using the older version of the CD-Bonn NN potential and the NCSM calculations
in basis spaces up to only $10\hbar\Omega$. Our binding energy results are tabulated 
in Tables \ref{tab_li6_he6} and \ref{tab_li6} with our conservative error estimates
based on the frequency and basis size sensitivity. 

\begin{figure}
\vspace*{2cm}
\includegraphics[width=7.0in]{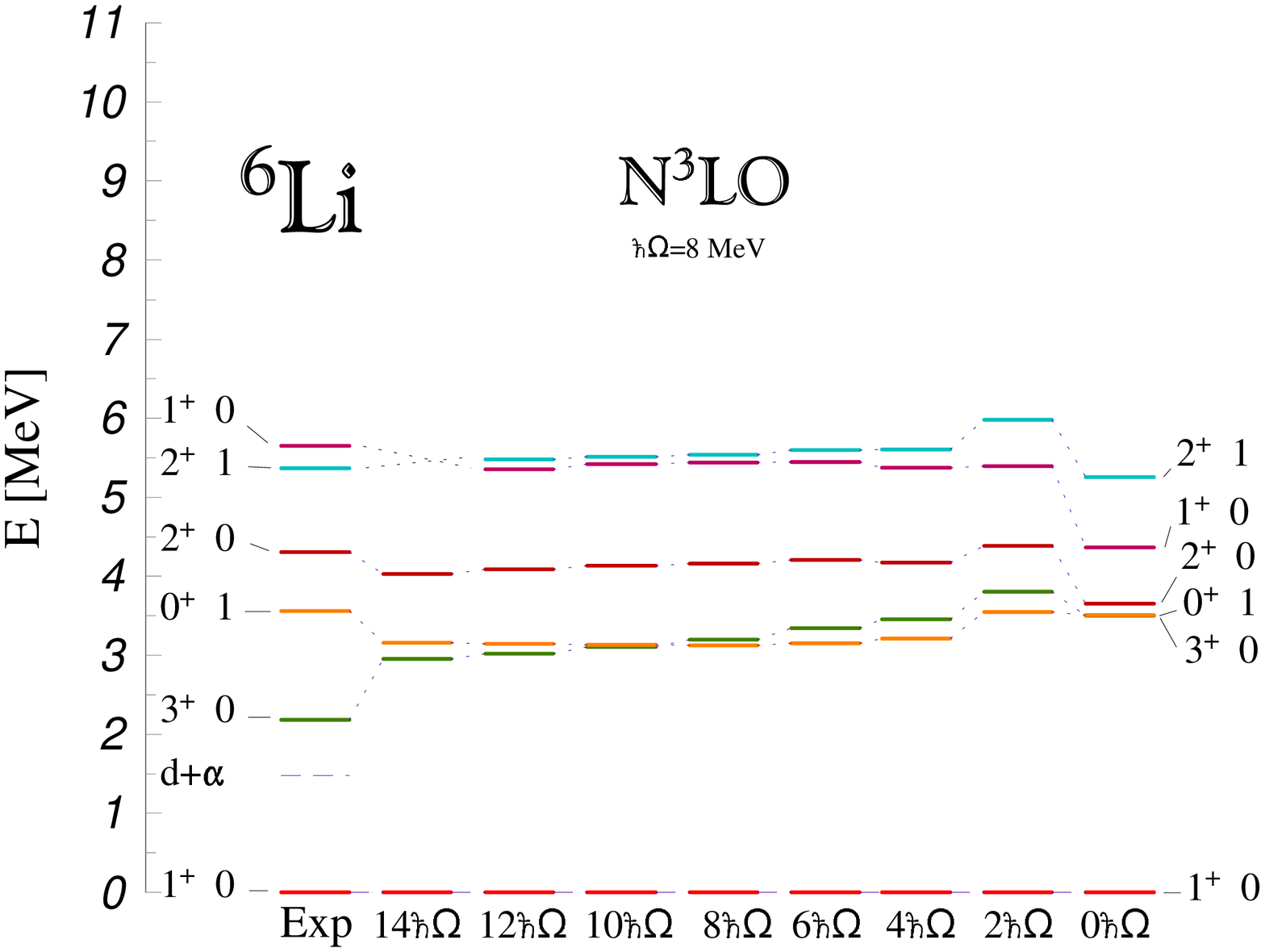}
\caption{\label{li6_exc_8} (Color online)
Calculated positive-parity excitation spectra of
$^{6}$Li obtained in $0\hbar\Omega$-$14\hbar\Omega$ 
($12\hbar\Omega$ for the $2^+ 1$ and $1^+_2 0$ states) 
basis spaces using two-body effective
interactions derived from the N$^3$LO NN potential
are compared to experiment. The HO frequency of $\hbar\Omega=8$ MeV was used.
The experimental values are from Ref. \protect\cite{AS88}.
}
\end{figure}

\begin{figure}
\vspace*{2cm}
\includegraphics[width=7.0in]{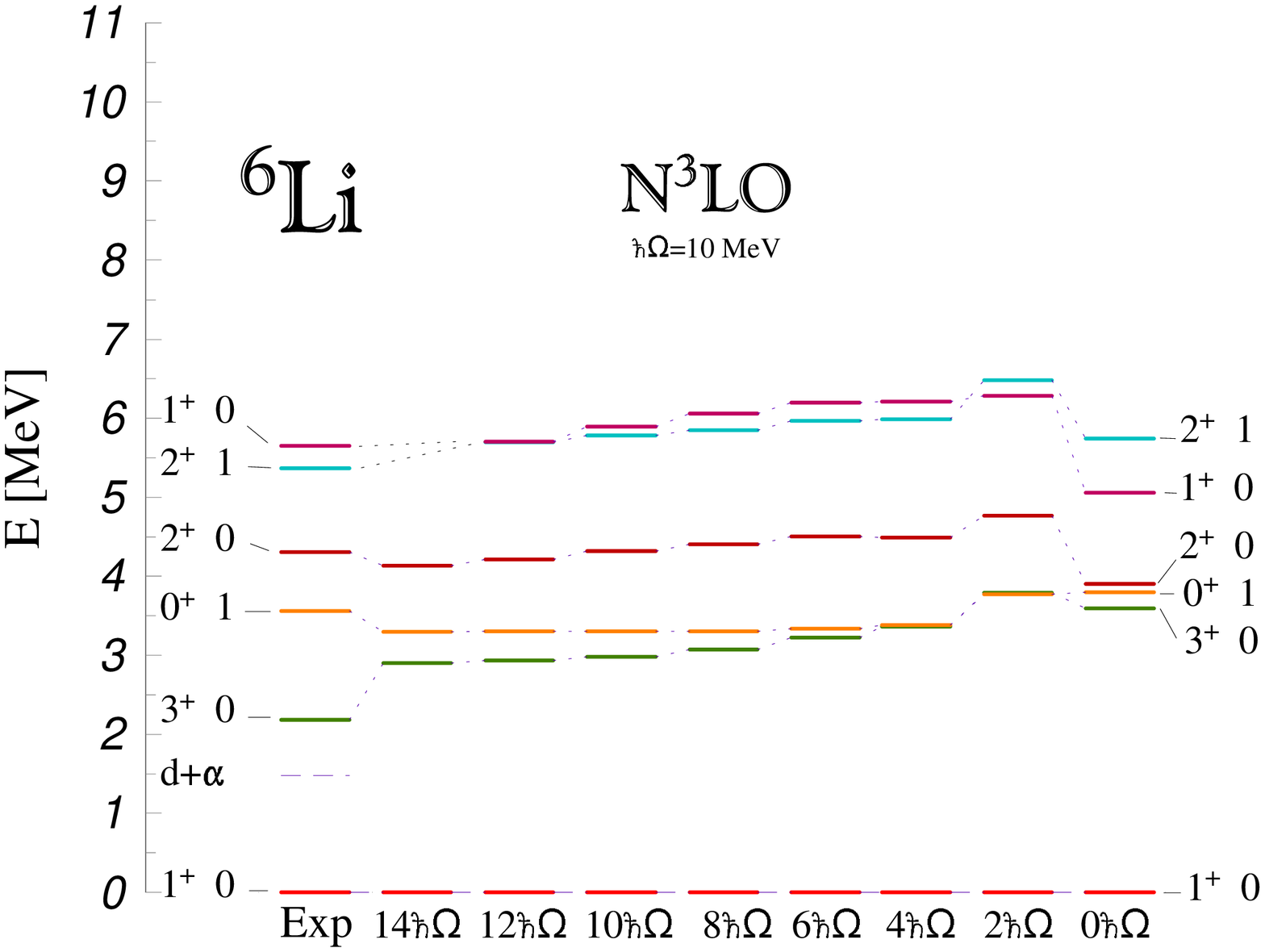}
\caption{\label{li6_exc_10} (Color online)
The same as in Fig. \protect\ref{li6_exc_8} for the HO frequency of $\hbar\Omega=10$ MeV.
}
\end{figure}

\begin{figure}
\vspace*{2cm}
\includegraphics[width=7.0in]{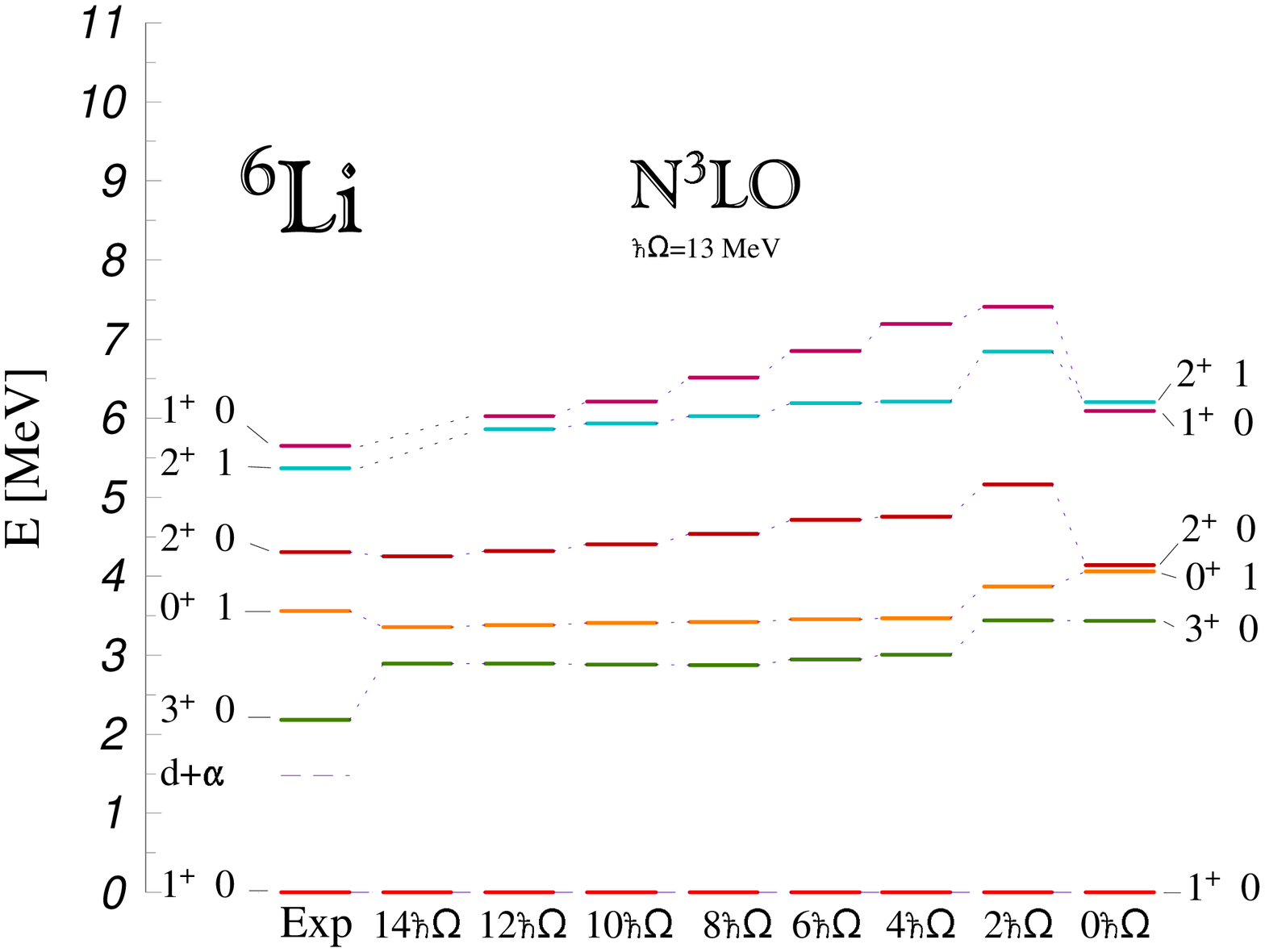}
\caption{\label{li6_exc_13} (Color online)
The same as in Fig. \protect\ref{li6_exc_8} for the HO frequency of $\hbar\Omega=13$ MeV.
}
\end{figure}

\begin{figure}
\vspace*{2cm}
\includegraphics[width=7.0in]{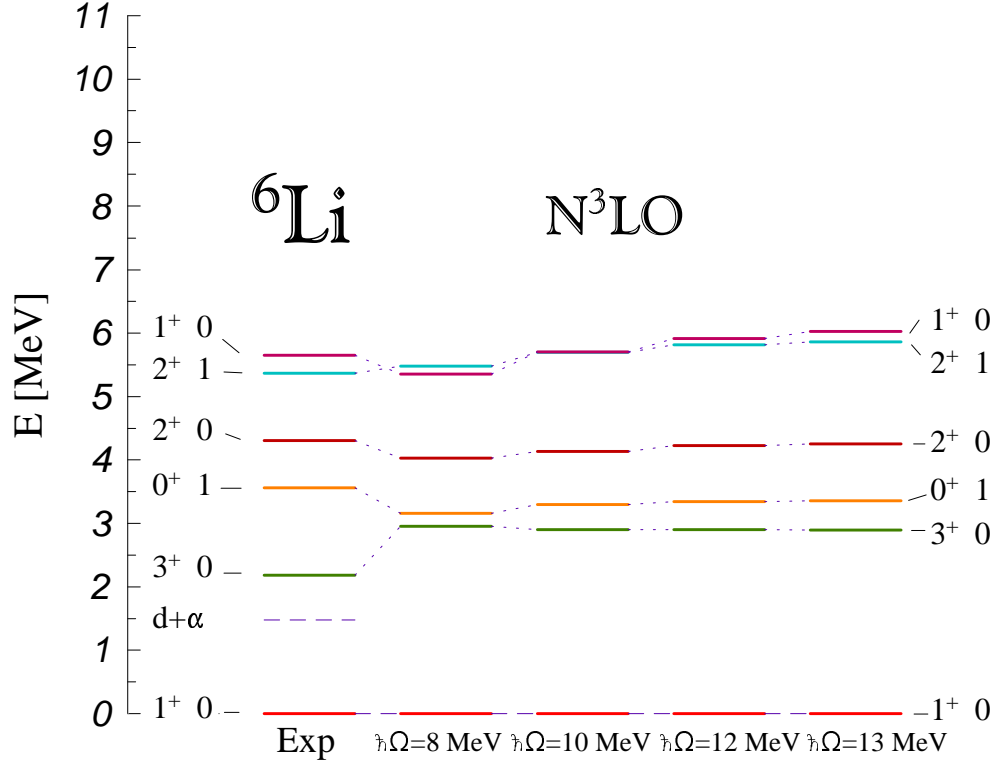}
\caption{\label{li6_exc_hw} (Color online)
Calculated positive-parity excitation spectra of
$^{6}$Li obtained in the $14\hbar\Omega$ ($12\hbar\Omega$ for the $2^+ 1$ and $1^+_2 0$ states) 
basis space using two-body effective interactions derived from the N$^3$LO NN potential
are compared to experiment. The HO frequency dependence in the range of 
$\hbar\Omega=8-13$ MeV is presented.
The experimental values are from Ref. \protect\cite{AS88}.
}
\end{figure}

We investigated the excitation energies of the five lowest excited states of $^6$Li
using the N$^3$LO NN potential. As the binding energy minimum shifts with the change of the basis size
and also as we expect that the energy of different excited states might have different dependence
on the HO frequency than the ground state, we calculated the excitation energies 
for several HO frequencies. Due to the complexity of the calculations, the lowest four state 
were obtained in basis spaces up to $N_{\rm max}=14$ while we stopped at $N_{\rm max}=12$
for the $2^+ 1$ and the $1^+_2 0$ state. The NCSM excitation energy dependence on the basis size
is presented in Figs. \ref{li6_exc_8}, \ref{li6_exc_10} 
and \ref{li6_exc_13} for the HO frequencies
of $\hbar\Omega=8, 10$ and 13 MeV, respectively. In Fig. \ref{li6_exc_hw}, we then show
the excitation energies for the $\hbar\Omega=8, 10, 12$ and 13 MeV HO frequencies 
in the largest basis space used.
It is apparent that the convergence rate with $N_{\rm max}$ is different for different states.
In particular, the $3^+ 0$ state and the $0^+ 1$ state converge faster in the higher frequency 
calculations, while the higher lying states converge faster in the lower frequency calculations.
As can be seen from Fig. \ref{li6_exc_hw} in the largest space used the dependence on the HO 
frequency is very weak. We summarize our excitation energies with the estimated errors
in Table \ref{tab_li6_he6}. We have, in particular, confidence in our $3^+ 0$ excitation
energy that is reflected in a very small associated uncertainty. In the $N_{\rm max}=14$
space, the $3^+ 0$ excitation energy is almost frequency independent in the range of 
$\hbar\Omega=10-13$ MeV. At $\hbar\Omega=8$ MeV, this state has a slightly higher excitation energy.
However, it is seen from Fig. \ref{li6_exc_8} that its energy is still decreasing with the basis
size enlargement. On the other hand, this state is fairly stable and converged with $N_{\rm max}$ 
in the $\hbar\Omega=10-13$ MeV calculations. 

In general, our calculated levels are in the correct
order when compared to experiment for the lowest four states. Concerning the $2^+ 1$ and 
the $1^+_2 0$ state, those are obtained at almost the same energy and the uncertainty of our
results is too large to make a definitive prediction as to their ordering.
When comparing our results to experiment we notice a striking discrepancy for the $3^+ 0$
state that calculated excitation energy is about 0.7 MeV higher than the experimental one.
Taking into account our confidence in convergence of this state, this is a serious
discrepancy in the prediction of the N$^3$LO NN potential. The remaining states, when considering
our estimated errors, are in fairly good agreement with experiment. We note, however,
a significant underestimation of the splitting between the $3^+ 0$ and the $2^+ 0$ states.
This splitting is influenced by the spin-orbit interaction strength.    

In Table \ref{tab_li6}, we compare our N$^3$LO excitation energy results with our NCSM
results obtained by the CD-Bonn 2000 and the AV8$^\prime$ NN potentials. 
In addition, we also present the GFMC
$^6$Li results obtained using the AV8$^\prime$. We note that we have not performed as detailed
excitation energy calculations using the CD-Bonn 2000 and the AV8$^\prime$ as we did for the N$^3$LO.
The presented NCSM CD-Bonn 2000 and the AV8$^\prime$ excitation energies were obtained 
in the $N_{\rm max}=14$ basis space using the HO frequency of $\hbar\Omega=12$ MeV and
$\hbar\Omega=11$ MeV, respectively. We expect, however, that the error estimates that we found
for our N$^3$LO excitation energy results are also applicable to our tabulated CD-Bonn 2000 
and AV8$^\prime$ excitation energy calculations. It is apparent that when considering
the estimated uncertainties our NCSM calculations cannot discriminate between the three
realistic NN potentials as to their predictions of the excitation energies. The CD-Bonn
gives a slightly lower excitation energy of the $3^+ 0$ state and a slightly larger
splitting between the $3^+ 0$ and the $2^+ 0$ states. However, the differences are smaller
than the difference between our NCSM AV8$^\prime$ and the GFMC AV8$^\prime$ results.
In general, the NCSM and the GFMC results using the AV8$^\prime$ are in a reasonable agreement. 
Although we have an about 200 keV difference in the $3^+ 0$ excitation energy, the difference with
respect to experiment is about 800 keV for the NCSM and about 1 MeV for the GFMC.
This further supports our statement about the failure of the accurate NN potentials
to predict the position of this lowest excited state of $^6$Li. 

The $^6$Li with the CD-Bonn and the AV8$^\prime$ was investigated within the NCSM
in Refs. \cite{NCSM_6,v3eff} using the basis spaces up to $N_{\rm max}=10$ with the two-body
effective interaction and $N_{\rm max}=6$ with the three-body effective interaction. 
The results that we present here are consistent with the previously published results.
Only the $2^+ 0$ excitation energy reported in this work is better converged and smaller.
This is due to the larger basis space, i.e. $N_{\rm max}=14$, employed here.   
We note that there is approximately a 500 keV difference in the excitation energy
of the $0^+ 1$ state obtained in the NCSM and the GFMC using the AV8$^\prime$.
It is interesting to point out that the NCSM calculations using the three-body effective
interaction in the $N_{\rm max}=6$ space 
predict this $0^+ 1$ state at a higher excitation energy closer to the GFMC result
than the NCSM calculations that employ just the two-body effective interaction 
in similar basis spaces \cite{v3eff}. This is contrary to the $T=0$ states
excitation energy results that are very similar using both the two-body or the three-body 
effective interaction \cite{v3eff}. The difference for the $0^+ 1$ state in the two NCSM
approximations needs further investigation. Ultimately, with the basis size enlargement
the results in both approximations must by construction converge to the same value.

In Table \ref{tab_li6_he6}, we also present electromagnetic properties of $^6$Li, 
binding energy of $^6$He and the Gamow-Teller $^6$He$\rightarrow ^6$Li B(GT) value
obtained in our NCSM calculations using the N$^3$LO NN potential. The $^6$He
binding energy is consistent with our $^6$Li results and under-binds experiment 
by about 3 MeV. The NCSM calculated electromagnetic properties, obtained 
using bare charges, are in a reasonable 
agreement with experiment. We note that while the B(M1) result is fairly stable
with the basis size change the B(E2) values increase in general as the basis size 
is increased. Consequently, our B(M1) values reported here are almost the same
as those published in Ref. \cite{NCSM_6} but our present B(E2) values are substantially 
larger. The present results were obtained using $N_{\rm max}=14$ while those of 
Ref. \cite{NCSM_6} using $N_{\rm max}=10$. We note that our B(GT) value 
for the $^6$He$\rightarrow ^6$Li ground-state to ground-state transition
over-predicts experiment. Recently, Schiavilla 
and Wiringa found that the AV18/Urbana-IX interaction also over-predicts the 
$^6$He$\rightarrow ^6$Li B(GT) value \cite{Schia}. In fact, their obtained Gamow-Teller
matrix elements ($\equiv\sqrt{{\rm B(GT)}}$), 2.254(5) and 2.246(10) using 
two types of wave functions, compare well with our result 2.28 obtained using the N$^3$LO
NN potential.

\begin{figure}
\vspace*{2cm}
\includegraphics[width=6.0in]{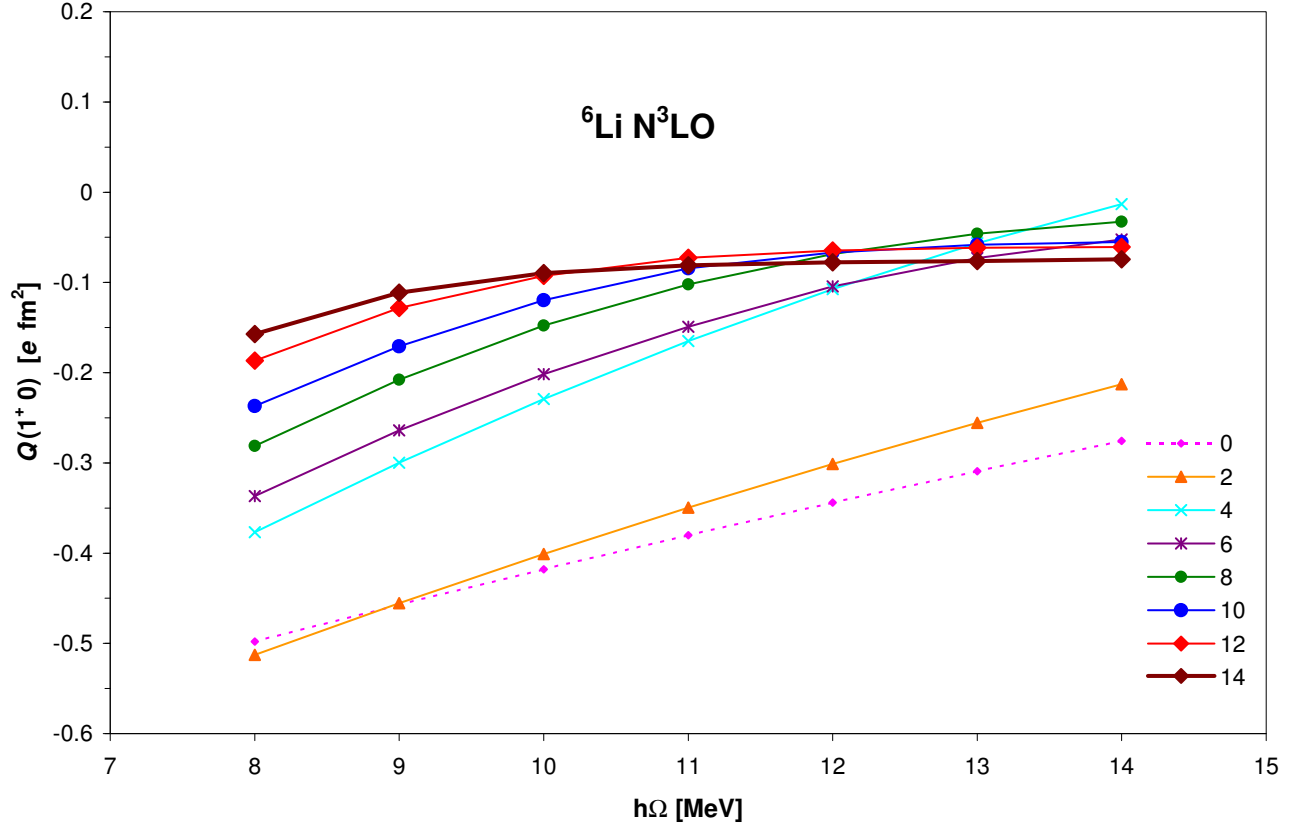}
\caption{\label{li6_n3lo_q} (Color online)
Harmonic-oscillator frequency dependence of the $^{6}$Li quadrupole moment
obtained in $0\hbar\Omega$-$14\hbar\Omega$ ($N_{\rm max}=0-14$) 
basis spaces using two-body effective interactions derived from
the N$^3$LO NN potential. 
}
\end{figure}

In order to judge the convergence of other observables than just energies,
in Fig. \ref{li6_n3lo_q} we present the HO frequency dependence of the $^6$Li ground-state
quadrupole moment for different basis sizes. Similar conclusions as for the $^4$He radius
calculations can be drawn here. With the basis size increase the frequency dependence
decreases. At the same time, the fastest convergence is not necessarily obtained for 
the same frequency as for the binding energy. A speed up of convergence for the
electromagnetic quadrupole observables should be achieved by using effective operators.
This has not been done here, but work on this problem is currently under way \cite{Ionel}. 
We conclude that the quadrupole moment of $^6$Li with the N$^3$LO 
interaction is -0.08(2) $e$ fm$^2$. 
 
To summarize this section, apart from more binding obtained using the CD-Bonn 2000
NN potential, we can see little differences in predictions of the $A=6$ nuclei
properties by the three accurate NN potentials, the N$^3$LO, the CD-Bonn 2000 
and the AV8$^\prime$. Let us just note that we also investigated 
the $2\hbar\Omega$-dominated (intruder)
states in $^6$He (not observed in experiment) using the three NN potentials.
Such states observed in heavier $p$-shell nuclei are in general 
predicted by the NCSM at too high energies and converge much 
more slowly than the $p$-shell dominated states \cite{Be8_NCSM,A10_NCSM}.
Using the same basis size and the same HO frequency, these states appear at
lower excitation energy when the N$^3$LO NN potential is used compared 
to the CD-Bonn 2000 and the AV8$^\prime$. This does not necessarily imply that
the N$^3$LO NN potential will predict the $2\hbar\Omega$-dominated 
states at lower energies but perhaps that
the NCSM convergence of such states is faster when the N$^3$LO NN potential is used.

\section{Application to $^{10}$B}\label{sec_res_b10}

In the previous section we pointed out that the N$^3$LO NN potential fails
to predict correct excitation energy of the $3^+ 0$ state in $^6$Li by about
700 keV. It has been realized that this problem which is common to other accurate
NN potentials is magnified in $^{10}$B. The ground state of $^{10}$B is $3^+ 0$
and the $1^+ 0$ state is the $^{10}$B first excited state at 0.72 MeV. 
Both the NCSM $^{10}$B calculations using the CD-Bonn and the Argonne
NN potentials and the GFMC $^{10}$B calculations using the Argonne NN potentials
show that these accurate NN potentials predict incorrectly
the $1^+ 0$ state as the ground state and the $3^+ 0$ state as an excited state
at higher than 1 MeV of excitation energy \cite{GFMC_9_10,v3eff,A10_NCSM}. 
This is a clear indication
for the need of a three-nucleon force not only to fix the under-binding problem but
also to correct, at least in some nuclei like $^{10}$B, the level ordering.
In fact, it has already been shown that, e.g., for the AV8$^\prime$ NN potential
the $^{10}$B ground-state spin problem is resolved by including certain
types of the three-nucleon forces like Illinois \cite{GFMC_9_10} or the Tucson-Melbourne
TM$^\prime$(99) \cite{NO03}, but it is not resolved by including 
the Urbana IX \cite{GFMC_9_10}. This type of sensitivity suggests that nuclear
structure calculations for light nuclei could be used to help determine the 
form and the parametrization of realistic three-nucleon forces that are unlike 
the two-nucleon interactions not well established.

\begin{figure}
\vspace*{2cm}
\includegraphics[width=7.0in]{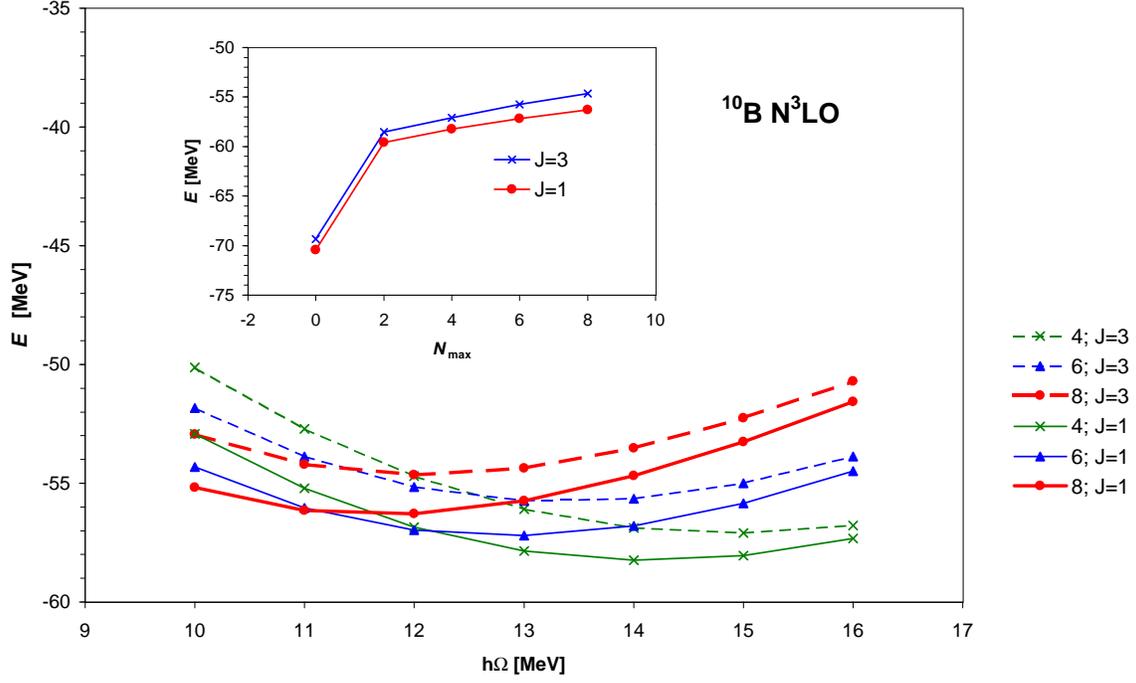}
\caption{\label{b10_n3lo_gs} (Color online)
The dependence of the $^{10}$B $1^+ 0$ and $3^+ 0$ state energy 
on the HO frequency for the $4\hbar\Omega$, $6\hbar\Omega$ and $8\hbar\Omega$ model
spaces calculated using the N$^3$LO NN potential. In the inset,
the $1^+ 0$ and $3^+ 0$ state energies at their respective HO frequency minima 
are plotted as a function of $N_{\rm max}$.
}
\end{figure}

\begin{figure}
\vspace*{2cm}
\includegraphics[width=7.0in]{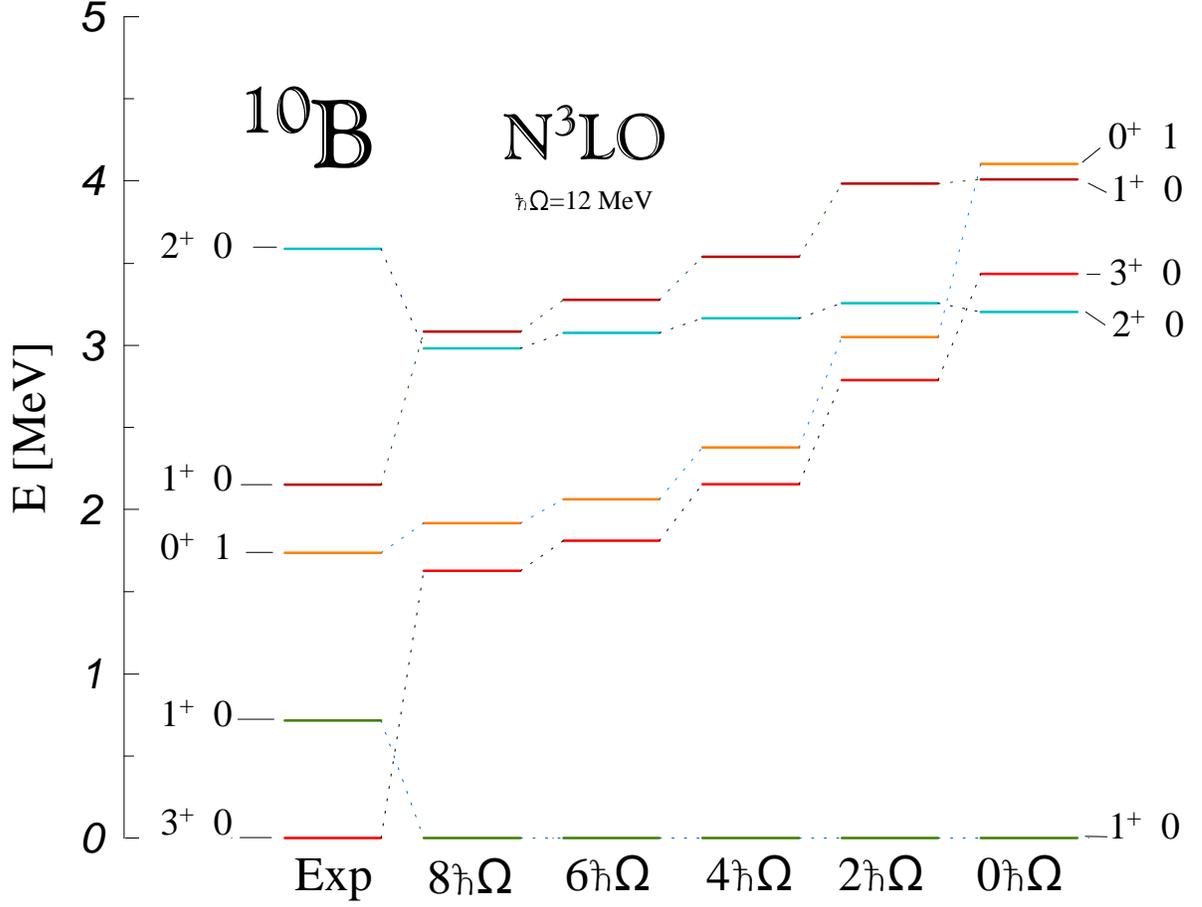}
\caption{\label{b10_n3lo_12} (Color online)
Calculated positive-parity excitation spectra of
$^{10}$B obtained in $0\hbar\Omega$-$8\hbar\Omega$ basis spaces using two-body effective
interactions derived from the N$^3$LO NN potential
are compared to experiment. The HO frequency of $\hbar\Omega=12$ MeV where the
$8\hbar\Omega$ ground state is at the minimum was used.
The experimental values are from Ref. \protect\cite{AS88}.
}
\end{figure}

It is quite interesting to perform the $^{10}$B calculations using the N$^3$LO
NN potential. Our NCSM results performed in basis spaces up to $N_{\rm max}=8$
are presented in Figs. \ref{b10_n3lo_gs} and \ref{b10_n3lo_12} and Table \ref{tab_b10}.
Unlike in the case of $^6$Li, we were not able to reach as large basis spaces 
that are needed for the excitation energy or the binding energy convergence. 
Still, our calculations
are sufficient to resolve the ground-state spin issue in $^{10}$B with the N$^3$LO
NN potential.    
In Fig. \ref{b10_n3lo_gs}, we present the dependence of the $^{10}$B $1^+ 0$ 
and $3^+ 0$ state energy on the HO frequency for different model spaces. 
In the NCSM, one typically determines the optimal HO frequency for the ground state
and uses the same frequency to describe the excited states. In general,
however, one should also investigate the the frequency dependence for the excited 
states in order to determine their energy \cite{v3eff}. This is in particular
important in the $^{10}$B investigation when we actually want to find out
which state is the ground state for a given NN potential.  
In the inset of Fig. \ref{b10_n3lo_gs}, the $1^+ 0$ and $3^+ 0$ state energies 
at their respective HO frequency minima are plotted 
as a function of $N_{\rm max}$. We can see that in the whole frequency range
that we show the $1^+ 0$ state is always the ground state. For the highest
frequency shown in the Figure, $\hbar\Omega=16$ MeV, the excitation energy
of the $3^+ 0$ state becomes quite small. However, we observe that with the basis
size increase the $3^+ 0$ excitation energy is actually increasing. 
In Fig. \ref{b10_n3lo_12}, we then show the excitation energies of the lowest 5 
$^{10}$B states calculated in the $0\hbar\Omega-8\hbar\Omega$ basis spaces
using a fixed HO frequency of $\hbar\Omega=12$ MeV. At this frequency, 
the ground-state energy is at the minimum in the $8\hbar\Omega$ space.
Clearly, the excitation spectrum is not converged as it was the case in our $^6$Li
calculations. However, there is definitely a trend to convergence. The differences
between the successive calculations become smaller and smaller with increasing 
$N_{\rm max}$. We can conclude with confidence that the $3^+ 0$ state will remain
the excited state with further increase of $N_{\rm max}$ and most likely will
appear at an excitation energy of more than 1 MeV. Based on our NCSM calculations, 
we conclude that the N$^3$LO NN potential fails to produce a correct ground-state 
spin for $^{10}$B.

We mentioned earlier that already at the N$^2$LO order of the chiral perturbation
theory the three-body terms appear. Those terms were not used in this work.
However, work on their inclusion in the NCSM is under way \cite{Andreas}. It is reasonable
to expect that these three-body terms will resolve the $^{10}$B ground state problem.
Structure of some of the EFT three-body terms in fact coincides 
with, e.g., the Tucson-Melbourne TM$^\prime$(99) three-nucleon interaction \cite{TMprime99}
that was already shown to resolve the $^{10}$B problem when combined with the
AV8$^\prime$ NN potential.

\section{Conclusions}\label{sec_concl}

We calculated the properties of $^3$H, $^4$He, $^6$Li, $^6$He and $^{10}$B
using the accurate charge-dependent
NN potential at fourth order of chiral perturbation theory.
We applied the {\it ab initio} no-core shell model and a variational
calculation in the harmonic oscillator basis. For $^6$Li, we were able to
reach the basis size of up to $16\hbar\Omega$ using an improved multi-shell 
(no-core) version of the shell model code ANTOINE. We obtained the $^6$Li
binding energy of 28.5(5) MeV and a converged excitation spectrum.
Properties of $^{10}$B were obtained using the same NN potential in 
basis spaces of up to $8\hbar\Omega$. Our results are consistent 
with results obtained by standard
accurate NN potentials and demonstrate a deficiency of Hamiltonians consisting
of only two-body terms. 
In addition to under binding compared to experiment for all investigated nuclei,
we found the $^6$Li $3^+ 0$ excitation energy 
over-predicted by about 700 keV compared to experiment. For $^{10}$B we found an 
incorrect ground state, $1^+ 0$, contrary to experimental $3^+ 0$.

We anticipate that most of these problems will be resolved by including
three-body terms that appear already at the third order of the chiral perturbation
theory. A work on including these terms in the NCSM is under way \cite{Andreas}. 

Finally, let us note that recently a new accurate non-local NN potential
has been constructed that fits not only the two-nucleon data but also
the binding energies of $^3$H and $^3$He without any need to introduce
a three-body interaction \cite{Doleschall}. 
As our calculations show, the deficiencies of two-nucleon
interactions are not limited only to the under binding but also to the 
nuclear structure issues most likely linked to an insufficient spin-orbit 
interaction strength. It will be very interesting to perform nuclear structure
calculations using the new non-local NN potential of Ref. \cite{Doleschall}
to test its predictions for the binding energies and also for the spin-orbit 
interaction strength sensitive observables.

\section{Acknowledgments}

We thank R. Machleidt for providing the N$^3$LO NN potential code.
We also thank A. Nogga for providing the $^4$He N$^3$LO binding energy result
obtained by the Faddeev-Yakubovsky method, K. Varga and W. Leidemann for the permission
to present their $^6$Li binding energy results for the Minnesota NN potential
prior to publication. 
This work was partly performed under the auspices of
the U. S. Department of Energy by the University of California,
Lawrence Livermore National Laboratory under contract
No. W-7405-Eng-48.

\begin{table}
\begin{tabular}{c|cccc}
       & NCSM$^a$     & Faddeev$^b$ & Yakubovsky$^c$ & Exp    \\
\hline
$^3$H  &  -7.85(1)    & -7.85       &    -           &  -8.48 \\
$^4$He & -25.36(4)    &   -         &  -25.41        & -28.30 \\
\end{tabular}

\noindent\small{$^a$ This work}

\noindent\small{$^b$ Ref. \protect\cite{Machl_priv}}

\noindent\small{$^c$ Ref. \protect\cite{Andreas}}
\caption{\label{tab_he4}
Calculated $^3$H and $^4$He ground state energies, in MeV, using the N$^3$LO
NN potential are compared to experiment. 
}
\end{table}

\begin{table}
\begin{tabular}{c|c|c}
 $^{6}$Li & Exp & N$^3$LO \\
\hline
$|E_{\rm gs}|(1^+ 0)$      &  31.995 &  28.5(5) \\
$Q_{\rm gs}$ [$e$ fm$^2$]        & -0.082(2) & -0.08(2) \\
$\mu_{\rm gs}$ [$\mu_{\rm N}^2$] & +0.822    & +0.839(5) \\
\hline
$E_{\rm x}(1^+_1 0)$ & 0.0    & 0.0  \\
$E_{\rm x}(3^+ 0)$   & 2.186  & 2.91(3) \\
$E_{\rm x}(0^+ 1)$   & 3.563  & 3.30(10) \\
$E_{\rm x}(2^+ 0)$   & 4.312  & 4.10(15) \\
$E_{\rm x}(2^+_1 1)$ & 5.366  & 5.50(30) \\
$E_{\rm x}(1^+_2 0)$ & 5.65   & 5.40(30) \\
\hline
B(E2;$1^+_1 0 \rightarrow 3^+ 0$)   & 21.8(4.8)  & 16.0(1.5)\\
B(M1;$0^+ 1 \rightarrow 1^+_1 0$)   & 15.42(32)  & 15.01(10) \\
B(E2;$2^+ 0 \rightarrow 1^+_1 0$)   & 4.41(2.27) & 6.2(8) \\
\hline
\hline
 $^{6}$He & Exp & N$^3$LO \\
\hline
$|E_{\rm gs}|(0^+ 1)$ & 29.269  & 26.2(5)\\ 
\hline
\hline
 $^{6}$He$\rightarrow ^6$Li & Exp & N$^3$LO \\
\hline
B(GT;$0^+_1 1\rightarrow 1^+_1 0$) & 4.728(15) & 5.22(10) \\
\end{tabular}
\caption{\label{tab_li6_he6} Experimental and calculated energies, in MeV,
quadrupole and magnetic moments,
as well as E2, in $e^2$ fm$^4$, and M1, in $\mu_N^2$, transitions
of $^{6}$Li and $^{6}$He as well as the B(GT)
values for the $^6$He ground state to $^6$Li ground state transition. 
Results obtained using the N$^3$LO NN potential are presented.
The errors are estimated from the HO frequency and basis size dependences
as well as the bare interaction result extrapolations.
The experimental values are from Ref. \protect\cite{AS88,Till02}.
}
\end{table}

\begin{table}
\begin{tabular}{c|c|c|c|cc|ccc}
                      &        &  NCSM    & NCSM      & NCSM         & GFMC         & NCSM     & SVM  & EIHH \\
                      & Exp    &  N$^3$LO & CD-Bonn   & AV8$^\prime$ & AV8$^\prime$ & MN       & MN   & MN   \\
\hline
$|E_{\rm gs}|(1^+ 0)$ & 31.995 &  28.5(5) & 29.35(40) & 28.5(5)      & 28.19(5)     & 36.50(4) & 36.51 & 36.64(7)\\
\hline
$E_{\rm x}(1^+ 0)$    & 0.0    & 0.0      &  0.0      &  0.0         &  0.0         &          &       &\\
$E_{\rm x}(3^+ 0)$    & 2.186  & 2.91(3)  &  2.86     &  2.97        &  3.21(7)     &          &       &\\
$E_{\rm x}(0^+ 1)$    & 3.563  & 3.30(10) &  3.29     &  3.41        &  3.94(8)     &          &       &\\
$E_{\rm x}(2^+ 0)$    & 4.312  & 4.10(15) &  4.47     &  4.28        &  4.10(6)     &          &       &\\
\end{tabular}
\caption{\label{tab_li6}
Experimental and calculated energies, in MeV, of $^6$Li. Binding and excitation state energy results 
for the realistic N$^3$LO, CD-Bonn 2000, AV8$^\prime$ 
NN potentials and the binding energy for the semi-realistic Minnesota (without Coulomb) NN potential
are presented. The GFMC \protect\cite{pieper01}, the SVM \protect\cite{Varga}
and the EIHH \protect\cite{Winfried} results are shown for comparison. See 
the text for further details.
}
\end{table}

\begin{table}
\begin{tabular}{c|cc}
$^{10}$B             & Exp    & N$^3$LO \\
\hline
$|E (3^+ 0)|$        & 64.751 & 54.65   \\  
$|E (1^+ 0)|$        & 64.033 & 56.28   \\
\hline
$E_{\rm x}(3^+_1 0)$ & 0.0    &  0.0    \\
$E_{\rm x}(1^+_1 0)$ & 0.718  & -1.628  \\
$E_{\rm x}(0^+_1 1)$ & 1.740  &  0.293  \\
$E_{\rm x}(1^+_2 0)$ & 2.154  &  1.459  \\
$E_{\rm x}(2^+_1 0)$ & 3.587  &  1.356  \\
\end{tabular}
\caption{\label{tab_b10} Experimental and calculated energies, in MeV,
of $^{10}$B. The NCSM results obtained using the two-body effective interaction
derived from the N$^3$LO NN potential in the $8\hbar\Omega$ basis space
and the HO frequency of $\hbar\Omega=12$ MeV.
The experimental values are from Ref. \protect\cite{AS88}
}
\end{table}

\end{document}